\documentclass[10pt,twocolumn,letterpaper]{article}

\usepackage{cvpr}              

\newcommand{\bx}{\mathbf{x}}
\newcommand{\bv}{\mathbf{v}}
\newcommand{\ba}{\mathbf{a}}
\newcommand{\bt}{\mathbf{t}}
\newcommand{\bz}{\mathbf{z}}
\newcommand{\by}{\mathbf{y}}
\newcommand{\bI}{\mathbf{I}}
\newcommand{\bp}{\mathbf{p}}
\newcommand{\bq}{\mathbf{q}}

\newcommand{\bb}{\mathbf{b}}
\newcommand{\bzero}{\mathbf{0}}
\newcommand{\bmu}{\boldsymbol{\mu}}
\newcommand{\bsigma}{\boldsymbol{\sigma}}
\newcommand{\bSigma}{\boldsymbol{\Sigma}}
\newcommand{\bepsilon}{\boldsymbol{\varepsilon}}
\newcommand{\bzeta}{\boldsymbol{\zeta}}

\renewcommand{\emph}[1]{#1}

\definecolor{groupgray}{RGB}{240,240,240}

\definecolor{cvprblue}{rgb}{0.21,0.49,0.74}
\usepackage[pagebackref,breaklinks,colorlinks,allcolors=cvprblue]{hyperref}

\title{3MDiT: Unified Tri-Modal Diffusion Transformer for Text-Driven \\ Synchronized Audio-Video Generation}

\author{
\textbf{Yaoru Li$^{1,2}$, }
\textbf{Heyu Si$^{1,2}$, }
\textbf{Federico Landi$^{2}$, }
\textbf{Pilar Oplustil Gallegos$^{2}$, }
\textbf{Ioannis Koutsoumpas$^{2}$, } \\
\textbf{O.~Ricardo Cortez Vazquez$^{2}$, }
\textbf{Ruiju Fu$^{2}$, }
\textbf{Qi Guo$^{2}\textsuperscript{*}$, }
\textbf{Xin Jin$^{2}$, }
\textbf{Shunyu Liu$^{3}\textsuperscript{*}$, }
\textbf{Mingli Song$^{1}$} \\
$^{1}$ Zhejiang University \qquad
$^{2}$ Huawei \qquad
$^{3}$ Nanyang Technological University \\
}

\usepackage{booktabs}
\usepackage{multirow}
\usepackage{amsmath}
\usepackage{makecell}
\usepackage{amssymb}
\usepackage{boxedminipage}
\usepackage{listings}
\usepackage{tabularx}
\usepackage{enumitem}
\usepackage{xcolor}
\usepackage{fontawesome5}
\usepackage{algorithm}      
\usepackage{algpseudocode}  
\usepackage{caption}
\usepackage{mathtools}
\usepackage{bm}
\usepackage[table]{xcolor}

\begin{document}
\maketitle

\begingroup
\renewcommand\thefootnote{}\footnote{
Work done during an internship at Huawei: liyaoru@zju.edu.cn. \\
*Correspondence to: guoqi39@huawei.com, shunyu.liu@ntu.edu.sg.
}
\addtocounter{footnote}{-1}
\endgroup

\begin{abstract}
Text-to-video~(T2V) diffusion models have recently achieved impressive visual quality, yet most systems still generate silent clips and treat audio as a secondary concern. Existing audio-video generation pipelines typically decompose the task into cascaded stages, which accumulate errors across modalities and are trained under separate objectives. Recent joint audio-video generators alleviate this issue but often rely on dual-tower architectures with ad-hoc cross-modal bridges and static, single-shot text conditioning, making it difficult to both reuse T2V backbones and to reason about how audio, video and language interact over time. To address these challenges, we propose \textbf{3MDiT}, a unified tri-modal diffusion transformer for text-driven synchronized audio-video generation.  Our framework models video, audio and text as jointly evolving streams: an isomorphic audio branch mirrors a T2V backbone, tri-modal omni-blocks perform feature-level fusion across the three modalities, and an optional dynamic text conditioning mechanism updates the text representation as audio and video evidence co-evolve. The design supports two regimes: training from scratch on audio-video data, and orthogonally adapting a pretrained T2V model without modifying its backbone. Experiments show that our approach generates high-quality videos and realistic audio while consistently improving audio-video synchronization and tri-modal alignment across a range of quantitative metrics.
\end{abstract}

\section{Introduction}
\label{sec:intro}

Pretrained text-to-video (T2V) diffusion models have recently become powerful enough to serve as foundation models for video generation~\cite{sora, wan, hunyuan, opensora, opensora2}. Large-scale systems such as Sora~\cite{sora} demonstrate that, with sufficient data and compute, a single model can render complex scenes with moving cameras, multi-actor interactions, and physically plausible dynamics over tens or even hundreds of frames. In parallel, open efforts including Wan~\cite{wan} and Hunyuan~\cite{hunyuan} show that a diffusion transformer~(DiT)~\cite{dit}-style spatiotemporal backbone trained on video latents is not only practical but also reusable: the same backbone can be prompted, tiled, and controlled across many text and layout conditions. Yet a key limitation remains widespread: T2V models only produce silent videos.

Silence is not just an aesthetic shortcoming. In human perception, audio and video form a coupled signal: wind should rise with moving treetops, and rainfall patter should coincide with visible raindrops and spreading ripples. Without synchronized audio, even visually strong videos appear synthetic or detached. This is why recent research has begun to shift towards text-to-audio-video~(T2AV) generation. On the one hand, video-to-audio (V2A) or foley-style models~\cite{syncfusion, sta, moviegen, thinksound, klingfoley, hunyuanvideofoley} synthesize sound based on videos, which can be generated by T2V methods; on the other hand, audio-to-video (A2V) methods~\cite{wan-s2v, avalign, avsync, mocha, MTVCraft} take audio as the driving signal and infer motion, camera, or scene dynamics that are consistent with it, which can be generated by text-to-audio~(T2A) methods. These pipelines are effective in narrow domains, but they are sequential, so noise and semantic mistakes propagate, and they are typically decoupled in training, so the two modalities do not co-evolve under a single alignment objective. The result is often small temporal misalignment or semantically correct but off-beat audio, especially in non-periodic or cinematic scenes. This has motivated joint audio-video diffusion models~\cite{mmdiffusion, seeingandhearing, avdit, mmldm, uniform, javisDiT, universe-1, ovi, bridgedit, svg}, which model both modalities in one transformer. The general recipe is appealing: represent video as spatiotemporal tokens, represent audio as temporally dense tokens, encode text as a conditioning sequence, and let a stack of multimodal DiT blocks discover synchronization. 

However, two challenges remain central to the design of joint audio-video generators: how to realize effective cross-modal interaction, and how to structure text conditioning. On the interaction side, most recent systems adopt a dual-tower architecture, where video and audio are modeled by separate DiT-style backbones and communication is introduced through additional architectural components. UniVerse-1~\cite{universe-1} performs block-wise stitching of pretrained experts to connect modality-specific towers. BridgeDiT~\cite{bridgedit} introduces bidirectional cross-attention bridges between T2V and T2A branches and partially unfreezes late layers to enable information flow. JavisDiT~\cite{javisDiT} incorporates prior-guided cross-modal attention and hierarchical priors tailored to its specific training pipeline. These designs demonstrate that explicit cross-modal coupling is beneficial, but they typically entangle fusion with the internals of the video backbone, making it harder to control fusion strength, to reuse different pretrained T2V models without architectural modifications, or to reason about how much of the visual prior is preserved. 

On the conditioning side, most joint models still rely on static single-shot text conditioning: a fixed prompt embedding is broadcast to both audio and video branches and remains unchanged throughout depth. This limits the model's ability to refine or re-weight textual cues as multimodal evidence emerges. BridgeDiT~\cite{bridgedit} takes an important step by decoupling captions into text–video and text–audio branches, which helps disentangle modality-specific supervision, but these captions are still injected as fixed signals that do not evolve along layers. As a result, fine-grained temporal cues and cross-modal dependencies must be inferred indirectly within each tower. We also note OmniFlow~\cite{omniflow}, a recent multi-modal rectified-flow model whose joint-attention design inspires part of our fusion strategy. However, OmniFlow is developed for static modalities such as images and audio, and therefore does not address the temporal alignment, motion dynamics, or cross-frame consistency required for audio-video generation. These observations motivate our design: a unified tri-modal framework in which video, audio, and text interact through omni-blocks, explicit fusion layers that can be cleanly added to existing T2V backbones, and an optional dynamic conditioning mechanism that allows the text representation itself to adapt as the audio and video streams co-evolve.

\begin{figure*}[htbp]
  \centering
  \includegraphics[width=1\linewidth]{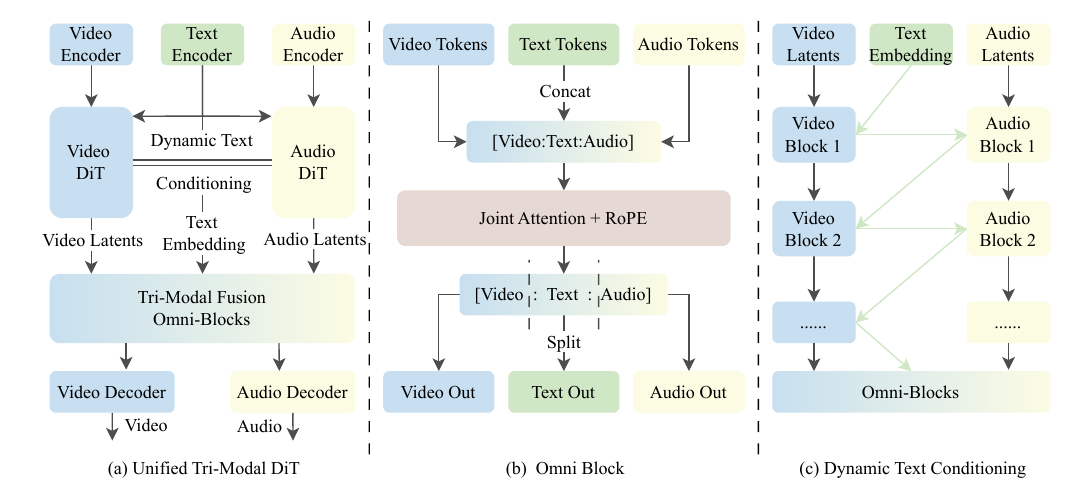}
  \caption{(a) Overall architecture: isomorphic Video/AudioDiT branches with pluggable omni-blocks; (b) Omni-blocks: concatenate $(x,y,a)$, apply 3D/1D RoPE to video/audio (text unrotated), perform joint attention, then split by modality and update with gated MLPs; (c) Dynamic text conditioning: between Video-DiT and Audio-DiT, a lightweight cross-attention dynamically refines the shared text representation to improve cross-modal synchrony.}
  \label{fig:dit}
\end{figure*}

To this end, we propose \textbf{3MDiT}, a novel framework centered on a unified tri-modal architecture as shown in Fig.~\ref{fig:dit}. First, we introduce omni-blocks inspired by OmniFlow~\cite{omniflow} that jointly aggregate video, text, and audio features and return modality-specific updates, which support both end-to-end training on audio-video data and minimally intrusive adaptation of strong T2V backbones. Second, we explore dynamic text conditioning: when the backbone exposes a text pathway, the text representation is treated as a state that can be updated as audio and video layers alternate, allowing evidence discovered in one modality to influence how the other interprets the prompt. In our framing, omni-blocks supply the general, backbone-agnostic mechanism for tri-modal fusion, while dynamic conditioning is a complementary option that further improves synchrony by supporting a mutable text stream.

Our contributions can be summarized as follows:
\begin{itemize}
\item We introduce \textbf{3MDiT}, a novel unified tri-modal diffusion transformer that can be trained from scratch on audio-video data or adapted on top of existing T2V backbones.
\item We design omni-blocks that merge video, text and audio at the feature level and can be attached orthogonally to a pretrained T2V backbone.
\item We explore a dynamic text conditioning mechanism that updates the text as the audio and video branches interact.
\item Our experiments and ablation studies demonstrate the effectiveness of our design and the promising synchronization and fidelity achieved by the proposed framework.
\end{itemize}

\section{Related Works}

\paragraph{Cascaded Audio-Video Generation.} To simplify the challenging task of text-to-audio-video (T2AV) generation, recent works explore cascaded pipelines that decompose the process into two stages. Two primary paradigms have emerged: (1) generating video from text first (T2V), then synthesizing synchronized audio conditioned on the generated video (V2A); and (2) generating audio from text first (T2A), followed by video generation conditioned on the synthesized audio (A2V). In particular, open-source foundation models for T2V~\cite{opensora, opensora2, wan, hunyuan} and T2A~\cite{audioldm, audioldm2} have achieved strong performance in generating high-quality videos and audios from text. Concurrently, methods for dubbing video content with sound (V2A) have made notable progress~\cite{syncfusion,sta,moviegen,klingfoley,hunyuanvideofoley,thinksound}, producing temporally aligned and realistic audio tracks. Similarly, approaches that generate video from speech or audio signals (A2V) have shown increasing capability in creating coherent visual content synchronized with input audio~\cite{avalign,avsync,mocha,MTVCraft,wan-s2v}. Although V2A and A2V models are typically developed independently, their architectures are theoretically applicable to such cascaded audio-visual generation paradigms.

\paragraph{Joint Audio-Video Generation.} While cascaded audio-video generation techniques have advanced significantly, synchronous T2AV generation remains challenging. Recent efforts focus on DiT-based architectures~\cite{sd3,opensora,flux1,videoldm} training on audio-video datasets~\cite{aist,aist++,landscape,mmtrail,TAVGBench,vggsound,vggss,avsync} for joint audio-video generation aiming to mitigate the noise accumulation. Seeing-Hearing~\cite{seeingandhearing} presents the task of T2AV, MM-Diffusion~\cite{mmdiffusion} introduces a diffusion block to facilitate cross-modal interaction, AV-DiT~\cite{avdit} leverages a frozen image-pretrained DiT backbone with lightweight adapters, MM-LDM~\cite{mmldm} designs a hierarchical multimodal autoencoder with modality-specific perceptual latent spaces, UniForm~\cite{uniform} unifies the generation paradigm through task tokens and a shared latent space, JavisDiT~\cite{javisDiT} introduces a hierarchical spatio-temporal prior estimator and an open-source benchmark, Universe-1~\cite{universe-1} and BridgeDiT~\cite{bridgedit} both fuse pre-trained T2V and T2A models, with the former adopting expert stitching and the latter introducing a dual-tower DiT and a visual-grounded captioning framework for disentangled text conditioning, Ovi~\cite{ovi} adopts symmetric twin DiTs for joint generation. In contrast to previous works, our approach is built around a unified tri-modal diffusion transformer that treats video, audio, and text as jointly modeled streams, rather than coupling two modality-specific towers via auxiliary bridges.

\section{Method}
\label{sec:method}
Our framework can be trained end-to-end from scratch.
When a pretrained T2V model is available, we instead treat the video tower as a frozen backbone and attach audio and fusion modules so that its visual prior is preserved during audio-video learning. Let $v \in \mathbb{R}^{B \times C_v \times F \times H \times W}$ denote video latents, where $B$ is the batch size, $C_v$ the latent channels, $F$ the number of frames, and $H \times W$ the spatial resolution. $a \in \mathbb{R}^{B \times L_a \times D}$ denotes audio latents, where $L_a$ is the sequence length. Text tokens are \(y \in \mathbb{R}^{B \times L_y \times D},\) where $L_y$ denotes the token length. Let $D$ be the shared dimension and $N_h$ the number of attention heads. We use $\sigma \in [0,1]$ as flow time and $t$ as the AdaLN~\cite{adaln} time embedding input. We use $(v,a,y)$ to denote model latents inside the DiT, and $(\tilde{v},\tilde{a},\tilde{y})$ to denote the generated outputs at inference time. 

\subsection{Preliminaries}
\label{sec:prelim}

We adopt Flow Matching (FM)~\cite{flowmatching}, a generative framework equivalent in law to DDPM~\cite{ddpm} that learns a continuous probability-flow ODE from noise to data. 

Our objective is to generate a synchronized audio-video pair $(\tilde{v},\tilde{a})$ from a text prompt $\tilde{y}$. For training, we draw clean targets \((v_{\mathrm{gt}}, a_{\mathrm{gt}})\) from the data distribution and independent standard Gaussian noise \((\epsilon_v, \epsilon_a)\) of the same shapes, \textit{i.e.}, \(\epsilon_v \sim \mathcal{N}(0,I)\), \(\epsilon_a \sim \mathcal{N}(0,I)\). For a sampled \(\sigma \sim p(\sigma)\) (\textit{e.g.}, uniform on \([0,1]\)), we form the linear interpolant
\begin{equation}
(v_\sigma, a_\sigma) \;=\; (1-\sigma)\,(v_{\mathrm{gt}}, a_{\mathrm{gt}}) \;+\; \sigma\,(\epsilon_v, \epsilon_a).
\end{equation}
The model predicts a velocity field \(u_\theta\) conditioned on text:
\begin{equation}
(\mathbf{u}_v, \mathbf{u}_a) \;=\; u_\theta\!\big((v_\sigma, a_\sigma), \sigma, y\big),
\end{equation}
and is trained to match the target velocity of rectified flow,
\begin{equation}
(\mathbf{u}_v^\star, \mathbf{u}_a^\star) \;=\; (\epsilon_v - v_{\mathrm{gt}},\; \epsilon_a - a_{\mathrm{gt}}).
\end{equation}
We minimize the weighted mean-squared error
\begin{equation}
\begin{aligned}
\mathcal{L}
= \mathbb{E}_{\sigma, v_{\mathrm{gt}}, a_{\mathrm{gt}}, \epsilon_v, \epsilon_a} \Big[
& w(\sigma)\,\big\| \mathbf{u}_v - (\epsilon_v - v_{\mathrm{gt}}) \big\|_2^2 \\
& {}+ w(\sigma)\,\big\| \mathbf{u}_a - (\epsilon_a - a_{\mathrm{gt}}) \big\|_2^2
\Big].
\end{aligned}
\end{equation}
Here \(w(\sigma)>0\) is a scalar weighting (\textit{e.g.}, \(w(\sigma)=1\) or \(w(\sigma)=\sigma(1-\sigma)\)). At inference, we integrate the probability-flow ODE
\(\tfrac{d}{d\sigma}(v_\sigma,a_\sigma)=u_\theta\big((v_\sigma,a_\sigma),\sigma,y\big)\)
from \(\sigma=1\) to \(\sigma=0\); classifier-free guidance~\cite{cfg} is applied in the conditional branch.

\subsection{Architecture}
\label{sec:arch}

Our architecture consists of three complementary components: an isomorphic audio branch, a dynamic text conditioning mechanism, and a set of omni-blocks. The audio branch mirrors the video backbone to enable audio generation. Dynamic text conditioning allows the text representation to evolve as audio and video layers alternate, enabling cross-branch information flow. Omni-blocks further couple the three modalities at deeper layers, providing explicit tri-modal fusion independent of the backbone structure.

\paragraph{Isomorphic Audio Branch.}
We introduce an isomorphic DiT branch for audio generation. We represent video as spatiotemporal tokens $v \in \mathbb{R}^{B \times L_v \times D}$ and audio as temporally dense tokens $a \in \mathbb{R}^{B \times L_a \times D}$, both conditioned on text tokens $y \in \mathbb{R}^{B \times L_y \times D}$. 
Here, $L_v$ denotes the tokenized video sequence length obtained after patchifying the $(F,H,W)$ latent into a sequence of spatiotemporal patches.
Generally, each DiT block follows a common pattern: (i) self-attention with rotary position embeddings on the active modality tokens (3D RoPE for video, 1D RoPE for audio), (ii) a per-block interaction between modality and text tokens, and (iii) an AdaLN-modulated gated MLP. We keep text tokens unrotated. The exact realization of the text interaction differs by backbone: Wan~\cite{wan} uses cross-attention from modality to text without updating the text stream, while Stable Diffusion 3~\cite{sd3} adopts a dual-stream joint attention that updates both the modality and the text.

\paragraph{Dynamic Text Conditioning.}
We introduce a dynamic text conditioning mechanism to enable text representations to evolve jointly with video and audio streams. Let the video and audio towers contain $L_{\mathrm{v}}$ and $L_{\mathrm{a}}$ DiT blocks, denoted $\{\mathcal{V}^{(i)}\}_{i=1}^{L_{\mathrm{v}}}$ and $\{\mathcal{A}^{(j)}\}_{j=1}^{L_{\mathrm{a}}}$. We keep hidden states $(v^{(\ell)}, a^{(\ell)}, y^{(\ell)})$ with initialization $(v^{(0)}, a^{(0)}, y^{(0)})$.

Static conditioning fixes the text stream and updates modalities separately:
\begin{align}
v^{(i)} \;&=\; \mathcal{V}^{(i)}\!\big(v^{(i-1)},\, y^{(0)},\, t(\sigma)\big),\\
a^{(j)} \;&=\; \mathcal{A}^{(j)}\!\big(a^{(j-1)},\, y^{(0)},\, t(\sigma)\big),
\end{align}
so no audio-video interaction occurs before later fusion.

Dynamic conditioning maintains a shared text state and alternates video and audio updates in a dual-stream DiT fashion. Let $L = L_v + L_a$ and an alternating schedule $\{b_\ell\}_{\ell=1}^{L}$ with 
$b_\ell \in \{\texttt{"V"}, \texttt{"A"}\}$, where \texttt{"V"} and \texttt{"A"} 
indicate that the $\ell$-th block is a video block or an audio block, respectively.
Define the counters
\[
i_v(\ell)=1+\#\{j \le \ell : b_j = \texttt{"V"}\},
\]
\[
i_a(\ell)=1+\#\{j \le \ell : b_j = \texttt{"A"}\}.
\]

Let $\mathcal{B}$ denote a dual-stream DiT block that runs joint attention over the concatenated sequence $[v;\,y]$ or $[a;\,y]$, applies rotary position embeddings only to the modality span, and writes gated residuals back to both the modality and text streams, followed by per-stream AdaLN–MLP updates.

When $b_\ell = \texttt{"V"}$, we update
\begin{equation}
\begin{aligned}
(v^{(\ell)},\, y^{(\ell)}) \;&=\;
\mathcal{B}^{v}_{\,i_v(\ell)}\!\big(v^{(\ell-1)},\, y^{(\ell-1)},\, t(\sigma)\big),\\
a^{(\ell)} \;&=\; a^{(\ell-1)},
\end{aligned}
\label{eq:interleave-video}
\end{equation}
and when $b_\ell = \texttt{"A"}$, we update
\begin{equation}
\begin{aligned}
(a^{(\ell)},\, y^{(\ell)}) \;&=\;
\mathcal{B}^{a}_{\,i_a(\ell)}\!\big(a^{(\ell-1)},\, y^{(\ell-1)},\, t(\sigma)\big),\\
v^{(\ell)} \;&=\; v^{(\ell-1)}.
\end{aligned}
\label{eq:interleave-audio}
\end{equation}

When the backbone exposes a read-only text stream (\textit{e.g.}, WAN-style), the shared-update path above is disabled; we keep $y$ fixed and rely on the tri-modal fusion layers for audio-video coupling.

\paragraph{Omni-Blocks.}
We insert $M$ tri-modal omni-blocks before the audio-video heads to allow 
video $v$, text $y$, and audio $a$ to interact at the same hidden width $D$. 
Each omni-block operates on the concatenated sequence
\begin{equation}
z = [\,v;\,y;\,a\,] \in \mathbb{R}^{B \times (L_v + L_y + L_a) \times D},
\label{eq:concat_sequence}
\end{equation}

Within an omni-block, we compute joint attention across all three modalities.  
Let
\begin{equation}
Q = z W_Q,\qquad K = z W_K,\qquad V = z W_V,
\label{eq:qkv_projection}
\end{equation}
where $W_Q, W_K, W_V \in \mathbb{R}^{D \times D}$ are learnable projection matrices 
and RoPE is applied separately to the video and audio spans 
(3D for video tokens, 1D for audio tokens, and none for text).
The joint-attention output is
\begin{equation}
H = \mathrm{softmax}\!\Big(
    \frac{\mathrm{RoPE}(Q)\,\mathrm{RoPE}(K)^{\top}}{\sqrt{D}}
\Big)\,V.
\label{eq:joint_attention}
\end{equation}
We then apply the output projection $W_O \in \mathbb{R}^{D \times D}$:
\begin{equation}
O = H W_O = [\,O_v;\,O_y;\,O_a\,],
\label{eq:output_projection}
\end{equation}
where $O_v$, $O_y$, and $O_a$ correspond to the video, text, and audio spans,
respectively.

The block returns modality-specific updates
\begin{equation}
v' = v + O_v,\qquad y' = y + O_y,\qquad a' = a + O_a,
\label{eq:modality_update}
\end{equation}
optionally scaled by learnable per-branch gates. Stacking such 
omni-blocks increases cross-modal alignment capacity while keeping the 
underlying video and audio towers unchanged, which ensures the visual prior of the base model is not disrupted while introducing the tri-modal fusion.

\paragraph{Orthogonality and Plug-in Adaptation.}
Given any pretrained T2V backbone composed of $L_v$ DiT blocks with frozen parameters $\Theta_{\mathrm{v}}^{\ast}$, our architecture extends it in a structurally orthogonal manner.  
We instantiate an audio branch that is isomorphic to the video branch. Each block preserves the same DiT micro-architecture of self-attention, cross-attention to text, and gated MLP, except that its spatiotemporal geometry is replaced by a 1D layout and 1D rotary position embeddings. This minimal change allows the model to handle audio as a temporal signal under the same denoising formulation as video.

To enable cross-modal interaction, we optionally insert $M$ omni-blocks before the audio-video head. When $M=0$ and text conditioning is independent across branches, the video pathway is mathematically identical to the base model, ensuring functional equivalence. When omni-blocks are enabled, their learnable gates control the strength of tri-modal fusion, gradually introducing synchronization while preserving the frozen backbone.

This design provides a plug-in recipe for upgrading any T2V diffusion transformer into an audio-–video generator: given a pretrained video backbone, one can attach an isomorphic 1D audio branch, add omni-blocks for cross-modal fusion, and optionally make use of the dynamic text conditioning mechanism. No retraining of video weights is required, and the resulting system inherits the visual priors and scalability of the base model while gaining synchronized audio generation capability.

\section{Experiments}
We conduct a series of experiments to evaluate both the generative quality and the adaptability of the proposed architecture. Our goals are threefold: Can the model achieve competitive audio-video generation quality when trained from scratch on limited audio-video data? (Sec.~\ref{exp:main})
(2) Can the framework be seamlessly adapted to different pretrained T2V backbones while largely preserving the base model’s original video generation capability? (Sec.~\ref{exp:main}) (3) Do the proposed omni-blocks and the dynamic text conditioning mechanism improve temporal synchronization and cross-modal semantic alignment? (Sec.~\ref{sec:ablation})

\subsection{Experimental Setup}

\paragraph{Datasets.}
We train and evaluate our framework on two audio-video datasets with distinct characteristics.
\textbf{Landscape}~\cite{landscape} contains natural scenes with diverse environmental soundscapes such as wind, rain, flowing water, and ambience.
It serves as our primary benchmark for open-domain alignment and includes 900 training clips and 100 test clips.
\textbf{AVSync15}~\cite{avsync} features a broad range of acoustic events and categories, covering diverse settings with variable resolutions and clip durations.
We use 1,350 videos for training and 150 for testing.
Detailed specifications regarding resolution, duration, and preprocessing are provided in the Appendix.

\paragraph{Training Regimes.}\label{training-regimes} We consider three training regimes:
\begin{itemize}
    \item \textbf{SD3-From-Scratch.}  
    We train 3MDiT entirely from scratch on NPUs without loading any pretrained T2V weights on a SD3 backbone. 
    \item \textbf{SD3-Adapted.}  
    We train 3MDiT on a pretrained 11B SD3 T2V backbone and keep all video weights frozen. 
    \item \textbf{Wan-Adapted.}  
    We train 3MDiT on the open-source Wan2.1-T2V-1.3B model~\cite{wan} on GPUs.
    This setting represents a fully open pipeline and demonstrates the plug-in nature of our design.
\end{itemize}
This triplet of settings allows us to systematically study from-scratch training, high-capacity backbone adaptation, and lightweight adaptation on an open base model.

\paragraph{Models.}
Across all settings we follow the modular design described in Sec.~\ref{sec:method}.
Text prompts are encoded with a transformer-based text encoder~\cite{t5}.
For the Wan-based experiments we use the T5-style encoder and video VAE provided in Wan~\cite{wan}, while for the pretrained 11B backbone we use its corresponding text encoder and video VAE to ensure compatibility.
Audio is encoded using DAC~\cite{dac}, a neural audio codec from which we extract latent features (before the model quantizes into codes) which are the 1D input sequence of the audio branch. Our training data is sampled at 16\,kHz, and we therefore use a pretrained DAC model trained at the same sampling rate to ensure compatibility with the input distribution. During inference, DAC’s decoder is used to reconstruct the waveform from the generated latent representation.

\paragraph{Metrics.}
We report standard metrics for visual quality, audio quality, and cross-modal alignment.
For video quality, we use Fréchet Video Distance (FVD)~\cite{fvd} computed between generated and reference videos.
For audio quality, we use Fréchet Audio Distance (FAD)~\cite{fad}, which compares embeddings of generated and real audio.
To evaluate semantic consistency, we follow prior work and compute text–video and text–audio alignment using pretrained vision–language and audio–text models such as CLIP~\cite{clip}, CLAP~\cite{clap} and CAVP reporting cosine similarity based scores.
To measure audio-video synchrony, we adopt an AV-Align metric~\cite{avalign} which assesses whether salient acoustic events align with the associated visual content.
We also include IB-style multimodal consistency scores~\cite{imagebind} where applicable to probe joint embedding alignment across modalities. Finally, we adopt the AVH-Score and Javis-Score metrics used in JavisBench~\cite{javisDiT}, which are used for evaluating audio-visual synchronization performance.
Together, these metrics provide a balanced evaluation of perceptual fidelity and cross-modal synchronization.

\paragraph{Evaluation.}
We mainly evaluate our models on Landscape~\cite{landscape} and AVSync15~\cite{avsync}, where each model is trained and evaluated on a single dataset without any cross-dataset mixing.
All generated samples are evaluated with 44.1\,kHz audio sampling rate, if the original data has a sampling rate of 16 kHz (\textit{e.g.}, Landscape), it is upsampled to 44.1 kHz during evaluation. 
For inference, we use CFG~\cite{cfg} and maintain the native spatial resolution of the original dataset or the corresponding pretrained backbone during inference.

\begin{figure}[htbp]
\centering
\includegraphics[width=1\linewidth]{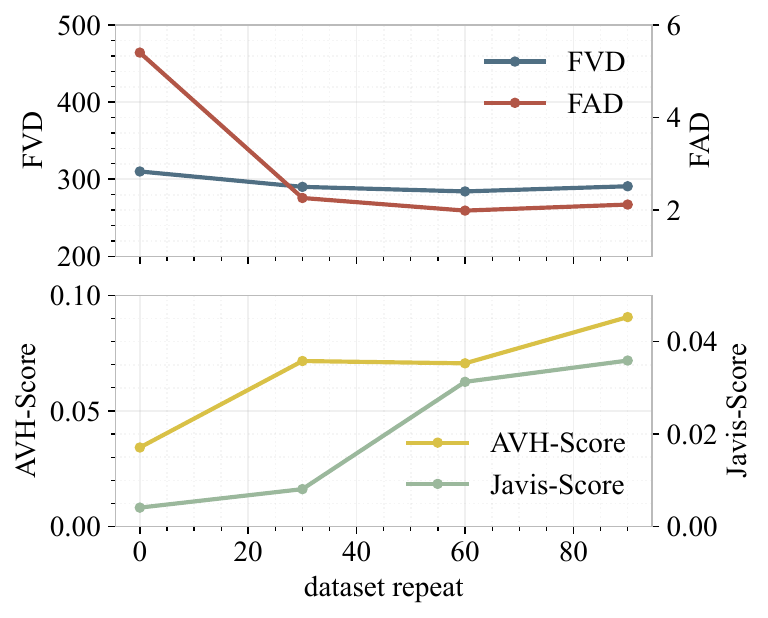}
\caption{
Metrics progress during training of the \textbf{Wan-adapted} model on the Landscape dataset.  
The horizontal axis indicates dataset repeats (number of full passes over the training set).  
Audio quality improves rapidly while visual fidelity remains stable, confirming compatibility between the frozen video backbone and learned audio pathways.
}
\label{fig:trend}
\end{figure}

\begin{figure*}[htbp]
  \centering
  \includegraphics[width=\textwidth]{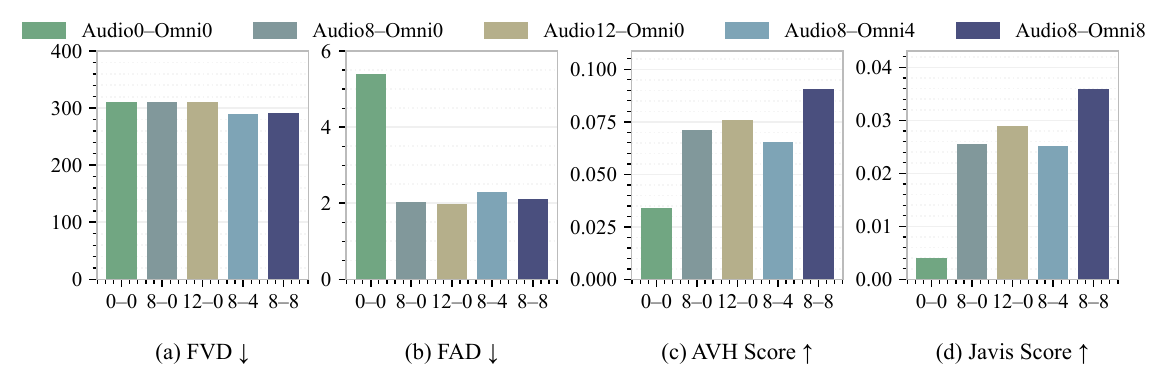}
  \caption{
    Effect of the number of audio blocks and omni-blocks on joint audio-video synchronization. ‘Audio8-Omni8’ denotes a model variant with 8 audio blocks and 8 omni-blocks; similarly, other naming conventions follow the same pattern.
    We report FVD/FAD (lower is better) and AVH/Javis scores (higher is better) under the \textbf{Landscape} dataset using the Wan~2.1 T2V-1.3B backbone as the base model, at 512$\times$288 and 10~fps with CFG=2. 
  }
  \label{fig:audio_omni_bars}
\end{figure*}

\begin{figure}[htbp]
  \centering
  \includegraphics[width=1\linewidth]{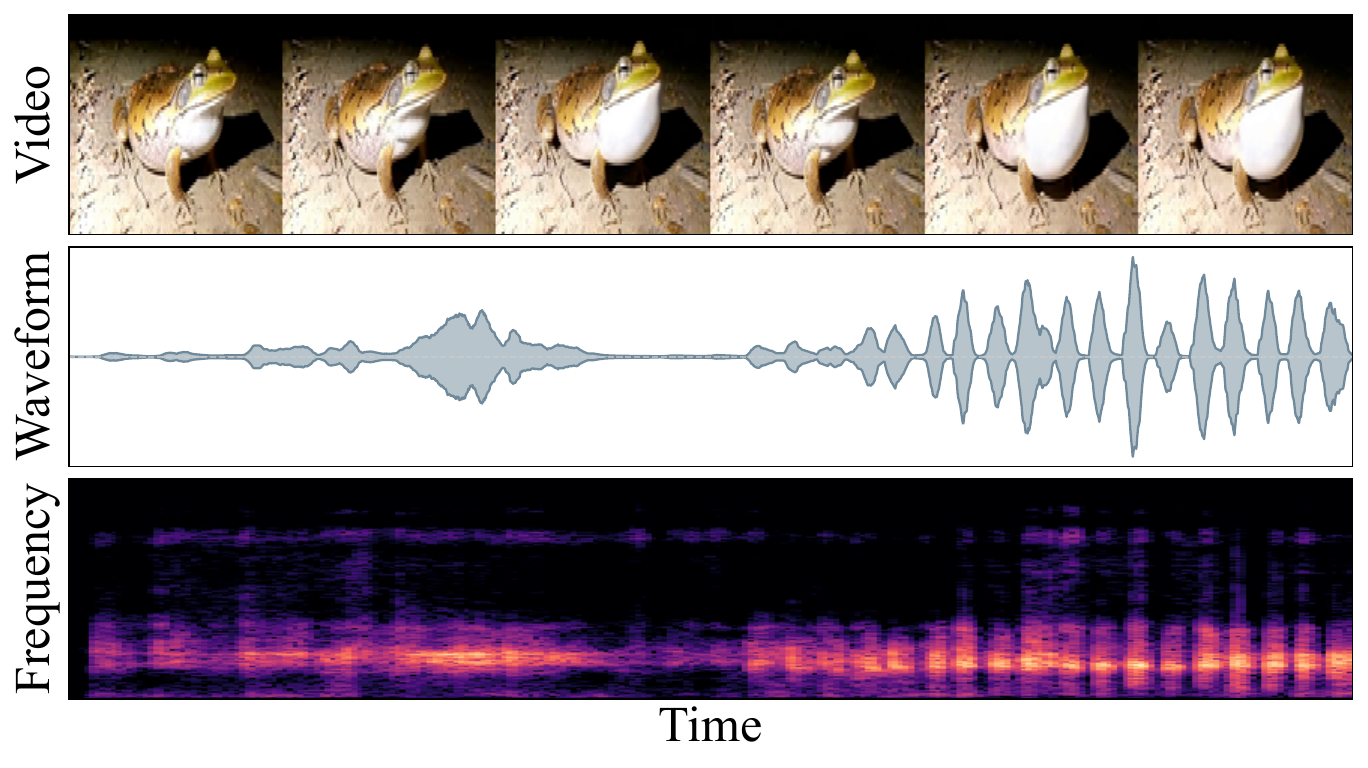}
  \caption{An example generated using  SD3 + A + D (referenced in Table~\ref{tab:avsync15}). Video (frames, on top) and audio (waveform and spectrogram) for the prompt: \textit{In a dimly lit, misty night scene, a large toad with mottled brown and white skin sits motionless on a patch of wet, sandy ground. (...) Audio: The deep, resonant croaking of a large toad echoes through the misty night air (...)}. The audio and video are aligned in time as the sound of the toad's croaking matches the movement of the toad's throat. The full caption specifies that while the throat moves, the mouth remains closed.}
  \label{fig:case_study}
\end{figure}

\begin{table*}[htbp]
\centering
\caption{
Evaluation on the \textbf{Landscape} test set. We integrate audio and video captions from~\cite{bridgedit} as text prompts. 
\textbf{Wan} denotes direct generation using the Wan 2.1 T2V-1.3B model;  \textbf{SD3} denotes direct generation using a SD3 T2V-11B model; 
\textbf{JavisDiT}~\cite{javisDiT} is evaluated following its original setup, generating 4-second 240p clips at 24 fps with CFG=7. 
\textbf{BridgeDiT}~\cite{bridgedit} results are taken from the original paper, which reports metrics on 5.4-second 480p clips at 15 fps. 
\textbf{3MDiT} denotes our method, refer to Sec.~\ref{training-regimes} for details.
All methods are evaluated with identical metric computation pipelines. 
All Wan and SD3 results correspond to 9.7-second clips at 10~fps with CFG=3 and a resolution of 512$\times$288, matching the native setting of the Landscape dataset.
Lower is better for FVD/FAD, higher is better for CLIP/CLAP/CAVP/IB-AV/AVAlign. Best results are highlighted in \textbf{bold} and second-best in \textit{italics}.
}
\label{tab:main}
\begin{tabularx}{\textwidth}{@{\extracolsep{\fill}}lcccccccc}
\toprule
\textbf{Model} & \textbf{FVD}$\downarrow$ & \textbf{FAD}$\downarrow$ &
\textbf{CLIP}$\uparrow$ & \textbf{CLAP}$\uparrow$ & 
\textbf{CAVP}$\uparrow$ & \textbf{IB-AV}$\uparrow$ & \textbf{AVAlign}$\uparrow$ \\
\midrule
\rowcolor{gray!15}\multicolumn{8}{l}{\textit{\textbf{T2V Base} (only video-related metrics)}} \\
Wan           & \textbf{282.8} & --     & 0.229 & --     & --     & --     & --     \\
SD3           & 709.0          & --     & 0.221 & --     & --     & --     & --     \\
\midrule
\rowcolor{gray!15}\multicolumn{8}{l}{\textit{\textbf{Previous Methods} (on different settings)}} \\
JavisDiT~\cite{javisDiT} & 408.6          & 5.55   & 0.226 & \textbf{0.265} & \textbf{0.795} & \textbf{0.208} & 0.098 \\
BridgeDiT~\cite{bridgedit} & 628.1        & 4.78   & --    & --     & --     & --     & 0.258 \\
\midrule
\rowcolor{gray!15}\multicolumn{8}{l}{\textit{\textbf{Our Method}}} \\
3MDiT (SD3-From-Scratch)  & 388.4 & 3.67 & 0.224 & 0.130 & 0.779 & 0.055 & \textit{0.626} \\
3MDiT (SD3-Adapted)   & 424.1 & \textbf{2.03} & 0.225 & \textit{0.185} & \textit{0.782} & \textit{0.068} & \textbf{0.627} \\
3MDiT (Wan-Adapted)   & \textit{283.1} & \textit{2.45} & \textbf{0.230} & 0.131 & 0.779 & 0.063 & 0.518 \\
\bottomrule
\end{tabularx}
\end{table*}

\begin{table*}[htbp]
\centering
\caption{
Ablation on the \textbf{AVSync15} test set using the SD3 11B backbone.  
To keep notation compact, ``SD3 Base'' denotes the frozen SD3 T2V base model;  
``A'' denotes audio blocks;  
``D'' denotes dynamic text conditioning;  
``O'' denotes omni-blocks.  
AVH and Javis denote the AVH-Score and Javis-Score from JavisBench~\cite{javisDiT}.  
Audio-related metrics for SD3 Base are computed from randomly generated noise and are reported only as a baseline reference. All models are evaluated with identical pipelines following the same settings as Table~\ref{tab:main}.  
Best results are highlighted in \textbf{bold} and second-best in \textit{italics}.
}
\label{tab:avsync15}
\begin{tabularx}{\textwidth}{@{\extracolsep{\fill}}lccccccccc}
\toprule
\textbf{Model} &
\textbf{FVD}$\downarrow$ & \textbf{FAD}$\downarrow$ &
\textbf{CLIP}$\uparrow$ & \textbf{CLAP}$\uparrow$ &
\textbf{CAVP}$\uparrow$ & \textbf{IB-AV}$\uparrow$ &
\textbf{AVH}$\uparrow$ & \textbf{Javis}$\uparrow$ &
\textbf{AVAlign}$\uparrow$ \\
\midrule
SD3 Base                & 1157.5 & 20.3 & 0.280 & 0.004 & --    & --    & --    & --    & --    \\
SD3 + A                 & 647.99 & 8.29 & 0.284 & 0.245 & \textbf{0.794} & 0.086 & 0.111 & 0.065 & 0.462 \\
SD3 + A + D             & \textbf{538.90} & \textbf{5.79} & \textbf{0.293} & \textit{0.306} & 0.792 & \textbf{0.190} & \textbf{0.166} & \textit{0.097} & \textbf{0.521} \\
SD3 + A + O + D         & \textit{573.14} & \textit{5.97} & \textbf{0.293} & \textbf{0.312} & \textbf{0.794} & \textit{0.181} & \textit{0.156} & \textbf{0.103} & \textit{0.500} \\
\bottomrule
\end{tabularx}
\end{table*}

\begin{table*}[htbp]
\centering
\caption{
Ablation on dynamic text conditioning under the SD3-style dual-stream architecture. 
Models are trained from scratch on the \textbf{Landscape} training split and evaluated on its held-out test split. 
The best results are highlighted in \textbf{bold}.
}
\label{tab:text_condition}
\begin{tabularx}{\textwidth}{@{\extracolsep{\fill}}lccccccccc}
\toprule
\textbf{Text Condition Method} &
\textbf{FVD}$\downarrow$ &
\textbf{FAD}$\downarrow$ &
\textbf{CLIP}$\uparrow$ &
\textbf{CLAP}$\uparrow$ &
\textbf{CAVP}$\uparrow$ &
\textbf{IB-AV}$\uparrow$ &
\textbf{AVAlign}$\uparrow$ &
\textbf{AVH}$\uparrow$ &
\textbf{Javis}$\uparrow$ \\
\midrule
Static Conditioning &
\textbf{288.6} & 2.45 &
\textbf{0.225} & 0.138 &
0.776 & 0.080 &
0.590 & 0.071 & 0.017 \\
Dynamic Conditioning &
383.1 & \textbf{1.86} &
0.220 & \textbf{0.213} &
\textbf{0.784} & \textbf{0.084} &
\textbf{0.615} & \textbf{0.078} & \textbf{0.050} \\
\bottomrule
\end{tabularx}
\end{table*}

\subsection{Main Experiments}

\paragraph{Quantitative Evaluation.}\label{exp:main} Table~\ref{tab:main} reports results from three groups of models: (i) T2V base backbones, where Wan~\cite{wan} and SD3~\cite{sd3} provide audio-free visual baselines; (ii) previous joint audio-video approaches such as JavisDiT~\cite{javisDiT} and BridgeDiT~\cite{bridgedit}, which are included for reference despite differences in training data, backbone design, and evaluation settings; and (iii) our 3MDiT variants, including 3MDiT (SD3-From-Scratch), 3MDiT (SD3-Adapted), and 3MDiT (Wan-Adapted). We further analyze several observations from Table~\ref{tab:main}. First, the SD3 T2V Adapted (11B) model achieves an FVD lower than its
T2V-only counterpart but higher than the SD3 Train-from-Scratch
variant.  
A likely explanation is that the pretrained 11B model exhibits a distribution
mismatch with the Landscape dataset, whereas training from scratch allows the
dual-stream backbone to fit the dataset more closely.  
Second, JavisDiT~\cite{javisDiT} shows strong semantic consistency between audio
and video (\textit{e.g.}, high IB-AV), which we attribute to its use of a considerably
larger training corpus (on the order of millions of samples) and shorter
4-second clips, making cross-modal alignment an easier objective.  In contrast, our evaluation uses longer 9.7-second clips without truncation or
other preprocessing, in line with our goal of supporting practical generation
settings.

The model adapted from the Wan backbone uses 8 audio blocks and 8 omni-blocks while keeping all 30 video blocks frozen.  
We track its training dynamics by evaluating checkpoints at different training steps.  
As shown in Fig.~\ref{fig:trend}, audio quality quickly improves from initial noise toward a well-structured distribution, whereas video quality stays highly stable.  
We further report the AVH-Score and Javis-Score~\cite{javisDiT}, both steadily improving as training progresses, indicating enhanced audio-video synchronization.

We acknowledge that fair cross-paper comparison is difficult because datasets, T2V priors, and inference settings often vary across works. Therefore, we explicitly describe the configurations and acknowledge that the reported numbers should not be interpreted as absolute differences.

\paragraph{Case Study.} Fig.~\ref{fig:case_study} presents a qualitative example.
The case includes video frames alongside the corresponding audio waveform and spectrogram.
These visualizations illustrate how our framework produces coherent audio-video pairs, with acoustic events that follow the temporal structure and semantics implied by the visual content.

\subsection{Ablation Studies}
\label{sec:ablation}

As a direct comparison with previous studies is limited by differences across methodologies, we want to emphasize internally consistent evaluations to examine the behavior of our framework under several configurations. To validate its effectiveness, generalizability, and robustness, we conduct comprehensive ablation studies.

\paragraph{Number of Audio and Omni-Blocks.}\label{exp:ablation-block} We evaluate multiple variants trained on the Landscape dataset using the Wan backbone. As shown in Fig.~\ref{fig:audio_omni_bars}, introducing audio blocks enables the model to learn audio generation from audio-video data while preserving the visual quality of the frozen T2V backbone. Adding omni-blocks consistently improves audio-video alignment, and does so without noticeably affecting either visual fidelity or acoustic quality. We further observe that increasing the number of audio blocks within a reasonable range leads to better audio quality, since a deeper audio DiT provides stronger modeling capacity for temporal structure and spectral detail. Moreover, adding more omni-blocks within a reasonable range also improves synchronization performance, as additional fusion layers give the model more opportunities to exchange information across modalities. Both observations indicate that the proposed architecture scales smoothly with capacity and provides a controllable way to balance quality and alignment.

\paragraph{Dynamic Text Conditioning.}\label{ablation-dynamic} To assess the contribution of dynamic text conditioning, we conduct ablation experiments under the SD3-style dual-stream architecture.  
The models in this study are trained from scratch, without loading any base model, on the Landscape training set, and evaluated on its held-out test split.  
We isolate the dynamic text update mechanism while keeping other configurations identical to the baseline with static text conditioning. Table~\ref{tab:text_condition} compares the two settings.  
Introducing dynamic text conditioning leads to noticeable improvement in temporal synchronization between the generated audio and visual streams, while maintaining both visual fidelity and acoustic quality within an acceptable variance range.  
This suggests that adaptive updates to the text representation can help the model capture early cross-modal cues, effectively encouraging tri-modal alignment.  
However, we also observe that the evolving text state may occasionally introduce small fluctuations in generation, indicating that excessive sensitivity to early textual changes could slightly perturb the dynamics of the diffusion process, and it is possible that sharing the text conditioning between DiTs is more useful for the audio reconstruction.
Overall, these findings highlight a trade-off between synchronization enhancement and stability, pointing to a promising direction for future refinement of dynamic conditioning mechanisms.

\paragraph{Generalization on AVSync15.}\label{ablation-avsync}
To evaluate the robustness of our framework under different data distributions, we further conduct controlled ablations on the AVSync15~\cite{avsync} dataset using the SD3 11B backbone.  
This dataset contains diverse acoustic scenes and variable resolutions and clip durations, making it a challenging testbed for audio-video synchronization.

We evaluate four variants:  
(i) the SD3 T2V base model (no audio pathway),  
(ii) adding only audio blocks,  
(iii) adding both audio blocks and omni-blocks, and  
(iv) enabling the full system with dynamic text conditioning.  
As shown in Table~\ref{tab:avsync15}, the variants trained on the AVSync15 dataset also exhibit promising audio-video generation performance.
The introduction of omni-blocks, together with the dynamic text conditioning mechanism based on the SD3 dual-stream architecture, leads to clear improvements in audio-video synchronization while maintaining stable visual and acoustic quality. However, adding omni-blocks alone does not yield a substantial gain.
We speculate that for a dataset with richer scenes and more complex acoustic–visual dynamics such as AVSync15, the additional trainable parameters introduced by omni-blocks may require longer training to fully realize their potential benefits.

\section{Conclusion}
\label{sec:conclusion}

We introduced 3MDiT, a unified tri-modal diffusion transformer for text-driven audio-video generation, where video, audio, and text are modeled as joint streams that interact throughout depth. By combining isomorphic audio blocks, plug-in omni-blocks, and an optional dynamic text conditioning mechanism, our framework allows the text representation and both modalities to co-evolve rather than relying on static captions or loosely coupled towers. Experiments on Landscape and AVSync15 show that this unified design yields high-quality audio and video while improving synchronization and cross-modal alignment across a range of metrics. We hope this tri-modal perspective provides a useful blueprint for future generative models that need to reason jointly over sound, motion, and language.

\small
\bibliographystyle{ieeenat_fullname}
\bibliography{main}


\clearpage
\appendix

\section{3MDiT Architecture}
\label{sec:app:arch}

This section details the implementation of 3MDiT used in our experiments,
complementing the high-level description in Sec.~\ref{sec:arch}.
We focus on (i) tokenization and positional embeddings,
(ii) six-slot AdaLN~\cite{adaln} modulation, (iii) SD-3~\cite{sd3} style dual-stream audio/video blocks,
(iv) tri-modal omni-blocks, and (v) conditioning and masking strategies.

\subsection{Tokenization and Positional Embeddings}

\paragraph{Video patchification.}
Let the input video latents be
\begin{equation}
v \in \mathbb{R}^{B \times C_v \times F \times H \times W},
\end{equation}
where $B$ is the batch size, $C_v$ the latent channels,
$F$ the number of frames, and $H \times W$ the spatial resolution.
We use a 2D patch embedder on each frame with patch size
$(p_h,p_w)$:
\begin{equation}
\label{eq:app:video-patch}
\Phi_{\mathrm{patch}}^{\mathrm{v}}: \mathbb{R}^{C_v \times H \times W}
\rightarrow \mathbb{R}^{L_s \times D},
\qquad
L_s = \frac{H}{p_h}\cdot\frac{W}{p_w},
\end{equation}
where $D$ is the hidden dimension.
Applying Eq.~\eqref{eq:app:video-patch} frame-wise yields a sequence
\begin{equation}
v^{(0)} \in \mathbb{R}^{B \times L_v \times D},
\qquad
L_v = F \cdot L_s,
\end{equation}
which is the input to the video DiT.

\paragraph{Audio embedding and 1D positional encoding.}
Let the audio latents be
\begin{equation}
a \in \mathbb{R}^{B \times L_a \times d_a},
\end{equation}
where $L_a$ is the audio sequence length and $d_a$ the codec latent dimension.
We first project to the model width $D$,
\begin{equation}
\label{eq:app:audio-linear}
\tilde{a}^{(0)} = a W_{\mathrm{a}} + \mathbf{b}_{\mathrm{a}},
\qquad
W_{\mathrm{a}} \in \mathbb{R}^{d_a \times D},
\ \mathbf{b}_{\mathrm{a}} \in \mathbb{R}^{D},
\end{equation}
and then apply a lightweight 1D convolutional position encoder
$\Psi_{\mathrm{pos}}^{\mathrm{a}}$ with residual connection:
\begin{equation}
\label{eq:app:audio-pos}
a^{(0)} = \tilde{a}^{(0)} + \Psi_{\mathrm{pos}}^{\mathrm{a}}\!\big(\tilde{a}^{(0)}\big),
\qquad
a^{(0)} \in \mathbb{R}^{B \times L_a \times D}.
\end{equation}
Here $\Psi_{\mathrm{pos}}^{\mathrm{a}}$ is implemented as a small stack of
depthwise 1D convolutions with SiLU activations.

\paragraph{3D RoPE for video and 1D RoPE for audio.}
We decompose the attention head dimension $d_{\mathrm{head}}$ into
temporal and spatial components for video:
\begin{equation}
d_{\mathrm{head}} = d_t + d_h + d_w,
\qquad
d_t : d_h : d_w \propto \rho_t : \rho_h : \rho_w,
\end{equation}
where $(\rho_t,\rho_h,\rho_w)$ is the rope-ratio used in the main model.
For a video token at frame index $f$ and spatial cell $(i,j)$,
the 3D RoPE angle vectors are
\begin{equation}
\label{eq:app:3d-rope}
\boldsymbol{\theta}_t(f) \in \mathbb{R}^{d_t},
\quad
\boldsymbol{\theta}_h(i) \in \mathbb{R}^{d_h},
\quad
\boldsymbol{\theta}_w(j) \in \mathbb{R}^{d_w},
\end{equation}
and are concatenated along the feature dimension before applying
the rotary embedding to the query and key subspaces.
For audio, we use a standard 1D RoPE over the time index $l$:
\begin{equation}
\label{eq:app:1d-rope}
\boldsymbol{\theta}_a(l) \in \mathbb{R}^{d_{\mathrm{head}}},
\end{equation}
and keep text tokens unrotated in all blocks.

\subsection{Six-Slot AdaLN Modulation}

Let $t(\sigma) \in \mathbb{R}^{D}$ be the time embedding at flow time
$\sigma \in [0,1]$, obtained from a sinusoidal encoder followed by an MLP.
For each modality $m \in \{v,y,a\}$,
we use a branch-specific MLP to produce a six-slot modulation vector
\begin{equation}
\label{eq:app:adaln-mlp}
T^{(m)}_6(\sigma) = \mathrm{MLP}^{(m)}_t\big(t(\sigma)\big)
\in \mathbb{R}^{6D}.
\end{equation}
We reshape and split $T^{(m)}_6$ into six $D$-dimensional slots:
\begin{equation}
\label{eq:app:adaln-split}
\big(
\Delta_{\mathrm{msa}}^{(m)},
\Gamma_{\mathrm{msa}}^{(m)},
g_{\mathrm{msa}}^{(m)},
\Delta_{\mathrm{mlp}}^{(m)},
\Gamma_{\mathrm{mlp}}^{(m)},
g_{\mathrm{mlp}}^{(m)}
\big)
\in \big(\mathbb{R}^{D}\big)^6,
\end{equation}
where
$\Delta_{\mathrm{msa}}^{(m)}$ and $\Gamma_{\mathrm{msa}}^{(m)}$
are the shift and scale for the self-attention branch,
$g_{\mathrm{msa}}^{(m)}$ is a scalar gate
(broadcast across the hidden dimension), and
$\Delta_{\mathrm{mlp}}^{(m)}$, $\Gamma_{\mathrm{mlp}}^{(m)}$,
$g_{\mathrm{mlp}}^{(m)}$ are the corresponding quantities for the MLP branch.

\paragraph{Modulated LayerNorm.}
Let $h^{(m)} \in \mathbb{R}^{B \times L_m \times D}$ be tokens of modality $m$
entering a block. We first apply a LayerNorm $\mathrm{LN}$ with no affine
parameters, then modulate it using the AdaLN~\cite{adaln} shift and scale:
\begin{equation}
\label{eq:app:adaln}
\widetilde{h}^{(m)}_{\mathrm{msa}}
=
\big(1 + \Gamma_{\mathrm{msa}}^{(m)}\big)
\odot \mathrm{LN}\!\big(h^{(m)}\big)
+ \Delta_{\mathrm{msa}}^{(m)},
\end{equation}
where $\odot$ denotes element-wise multiplication with broadcasting over
the sequence and batch dimensions. An analogous transformation
$\widetilde{h}^{(m)}_{\mathrm{mlp}}$ is defined using
$\Delta_{\mathrm{mlp}}^{(m)}$ and
$\Gamma_{\mathrm{mlp}}^{(m)}$.

\paragraph{Gated residual paths.}
Given an attention or MLP update $\Delta h^{(m)}$, the residual connection
is controlled by the learned gate:
\begin{equation}
\label{eq:app:gated-residual}
h^{(m)}_{\mathrm{out}} = h^{(m)}_{\mathrm{in}} +
g^{(m)} \odot \Delta h^{(m)},
\end{equation}
where $g^{(m)}$ is either $g_{\mathrm{msa}}^{(m)}$ or
$g_{\mathrm{mlp}}^{(m)}$, broadcast to the token shape.
When $g^{(m)}$ is initialized near zero, the residual branch starts close to
identity and can gradually learn stronger cross-modal interactions.

\subsection{SD-3 Style Dual-Stream DiT Blocks}

Each dual-stream block processes one modality (video or audio) and the text
stream jointly. Let
\begin{equation}
v \in \mathbb{R}^{B \times L_v \times D},
\quad
a \in \mathbb{R}^{B \times L_a \times D},
\quad
y \in \mathbb{R}^{B \times L_y \times D}
\end{equation}
denote the current hidden states.
We describe a generic block acting on modality tokens $x$
(either $v$ or $a$) and text $y$.

\paragraph{Joint attention over \texorpdfstring{$[x;\,y]$}{[x;y]}.}
We form the concatenated sequence
\begin{equation}
\label{eq:app:concat-xy}
z = [\,x;\,y\,] \in \mathbb{R}^{B \times (L_x + L_y) \times D}.
\end{equation}
Queries, keys, and values are obtained by three linear projections
followed by rotary position embeddings on the modality span:
\begin{equation}
\label{eq:app:dit-qkv}
\begin{split}
Q &= \mathrm{RoPE}_x\big(z W_Q\big),\\
K &= \mathrm{RoPE}_x\big(z W_K\big),\\
V &= z W_V,
\end{split}
\end{equation}
where $W_Q, W_K, W_V \in \mathbb{R}^{D \times D}$.
For video blocks, $\mathrm{RoPE}_x$ is the 3D rotary embedding
of Eq.~\eqref{eq:app:3d-rope}; for audio blocks, it is 1D RoPE
of Eq.~\eqref{eq:app:1d-rope}; and text positions are left unrotated.

The joint attention output is
\begin{equation}
\label{eq:app:dit-attn}
H = \mathrm{softmax}\!\Big(
    \frac{Q K^\top}{\sqrt{D}}
\Big)V,
\qquad
O = H W_O,
\end{equation}
with $W_O \in \mathbb{R}^{D \times D}$.
We split $O$ back into modality and text spans
\begin{equation}
O = [\,O_x;\,O_y\,],
\qquad
O_x \in \mathbb{R}^{B \times L_x \times D},
\quad
O_y \in \mathbb{R}^{B \times L_y \times D}.
\end{equation}

\paragraph{Per-stream AdaLN-MLP.}
Using the AdaLN parameters in
Eq.~\eqref{eq:app:adaln-split}, we obtain
modulated inputs for the attention and MLP branches:
\begin{equation}
\begin{split}
\widehat{x}_{\mathrm{msa}} &= \widetilde{x}^{\mathrm{(msa)}}, \quad
\widehat{y}_{\mathrm{msa}} = \widetilde{y}^{\mathrm{(msa)}}, \\
\widehat{x}_{\mathrm{mlp}} &= \widetilde{x}^{\mathrm{(mlp)}}, \quad
\widehat{y}_{\mathrm{mlp}} = \widetilde{y}^{\mathrm{(mlp)}},
\end{split}
\end{equation}
where the tilde denotes Eq.~\eqref{eq:app:adaln}.
The final updates are
\begin{equation}
\label{eq:app:dual-update}
\begin{split}
x' &= x + g_{\mathrm{msa}}^{(x)} \odot O_x
      + g_{\mathrm{mlp}}^{(x)} \odot \mathrm{MLP}_x(\widehat{x}_{\mathrm{mlp}}),\\
y' &= y + g_{\mathrm{msa}}^{(\mathrm{y})} \odot O_y
      + g_{\mathrm{mlp}}^{(\mathrm{y})} \odot \mathrm{MLP}_y(\widehat{y}_{\mathrm{mlp}}),
\end{split}
\end{equation}
where $\mathrm{MLP}_x$ and $\mathrm{MLP}_y$ are standard
two-layer feed-forward networks with GELU activations.

\paragraph{Static vs.\ dynamic text conditioning.}
As discussed in Sec.~\ref{sec:arch}, we support two execution modes:
(i) static text conditioning, where the text stream $y$ is kept fixed
as $y^{(0)}$ for both towers, and (ii) dynamic shared text conditioning,
where $y$ is updated as in Eqs.~\eqref{eq:interleave-video}
and~\eqref{eq:interleave-audio} by alternating video and audio blocks.
Both modes share the same block implementation; they differ only in whether
the text state $y^{(\ell)}$ is shared across towers.

\subsection{Wan-Style DiT Blocks}
\label{sec:app:wan-blocks}

For completeness, we also describe the Wan-style DiT blocks used when
adapting a pretrained WAN backbone~\cite{wan}. In this setting, the video
(and optionally audio) towers follow the original WAN design: each block
updates only the modality stream with self-attention, cross-attention to
text, and an AdaLN–MLP stack, while the text tokens remain read-only and
are never updated inside the backbone. As a result, this block does not
support our dynamic text conditioning mechanism.

\paragraph{Single-stream DiT with read-only text.}
Let $x \in \mathbb{R}^{B \times L_x \times D}$ denote the tokens of a
single modality (video or audio) entering a Wan-style block, and
$y \in \mathbb{R}^{B \times L_y \times D}$ the text tokens, which are
fixed for all layers (i.e., $y^{(\ell)} \equiv y^{(0)}$).
Time conditioning uses a six-slot modulation vector
\begin{equation}
T_6 = \mathrm{MLP}_t\big(t(\sigma)\big)
\in \mathbb{R}^{B \times 6 \times D},
\end{equation}
where $t(\sigma) \in \mathbb{R}^{B \times D}$ is the time embedding.
We split $T_6$ along the slot dimension:
\begin{equation}
\label{eq:app:wan-split}
\big(
\Delta_{\mathrm{msa}},\Gamma_{\mathrm{msa}},g_{\mathrm{msa}},
\Delta_{\mathrm{mlp}},\Gamma_{\mathrm{mlp}},g_{\mathrm{mlp}}
\big)
\in \big(\mathbb{R}^{B \times 1 \times D}\big)^6.
\end{equation}
Here, $\Delta_{\mathrm{msa}}$ and $\Gamma_{\mathrm{msa}}$ are the
shift and scale for the self-attention branch, $g_{\mathrm{msa}}$
is the attention gate, and
$\Delta_{\mathrm{mlp}},\Gamma_{\mathrm{mlp}},g_{\mathrm{mlp}}$
are the corresponding quantities for the MLP branch.

We define a modulation operator acting on LayerNorm outputs
\begin{equation}
\label{eq:app:wan-mod}
\operatorname{Mod}(u;\Delta,\Gamma)
=
\big(1+\Gamma\big)\odot \mathrm{LN}(u) + \Delta,
\end{equation}
with broadcasting over the batch and sequence dimensions.

\paragraph{Self-attention with RoPE.}
Given $x$ and precomputed rotary frequencies $\mathrm{RoPE}$ for
the modality (3D RoPE for video, 1D RoPE for audio),
the self-attention sub-layer proceeds as
\begin{equation}
\label{eq:app:wan-self}
\begin{aligned}
x_{\mathrm{msa}} &= \operatorname{Mod}\big(x;\Delta_{\mathrm{msa}},\Gamma_{\mathrm{msa}}\big),\\
Q &= \mathrm{RoPE}\!\big(x_{\mathrm{msa}} W_Q\big),
K = \mathrm{RoPE}\!\big(x_{\mathrm{msa}} W_K\big),
V = x_{\mathrm{msa}} W_V,\\[0.2em]
H_{\mathrm{msa}} &= \mathrm{softmax}\!\Big(\frac{QK^\top}{\sqrt{D}}\Big)V,\\
\widetilde{x} &= x + g_{\mathrm{msa}} \odot \big(H_{\mathrm{msa}} W_O\big),
\end{aligned}
\end{equation}
where $W_Q,W_K,W_V,W_O \in \mathbb{R}^{D \times D}$ are learned
projections and the gate $g_{\mathrm{msa}} \in \mathbb{R}^{B \times 1 \times D}$
is broadcast to the token shape.

\paragraph{Cross-attention to fixed text.}
WAN-style blocks then apply a cross-attention from $x$ to the
(read-only) text context $y$, without updating the text stream:
\begin{equation}
\label{eq:app:wan-cross}
\begin{aligned}
Q_{\times} &= \mathrm{LN}_q(\widetilde{x}) W^{\times}_Q,\quad
K_{\times} = \mathrm{LN}_k(y) W^{\times}_K,\quad
V_{\times} = y W^{\times}_V,\\[0.2em]
H_{\times} &= \mathrm{softmax}\!\Big(
    \frac{Q_{\times}K_{\times}^\top}{\sqrt{D}}
\Big)V_{\times},\\
\widehat{x} &= \widetilde{x} + H_{\times} W^{\times}_O,
\end{aligned}
\end{equation}
where the LayerNorms $\mathrm{LN}_q,\mathrm{LN}_k$ and
projections $W^{\times}_Q,W^{\times}_K,W^{\times}_V,W^{\times}_O$
follow the original WAN implementation. Note that $y$ is used
only as a key/value context and is not modified.

\paragraph{AdaLN-MLP update.}
Finally, an AdaLN–MLP branch refines the modality stream:
\begin{equation}
\label{eq:app:wan-mlp}
\begin{aligned}
x_{\mathrm{mlp}} &= \operatorname{Mod}\big(\widehat{x};\Delta_{\mathrm{mlp}},\Gamma_{\mathrm{mlp}}\big),\\
H_{\mathrm{mlp}} &= \mathrm{MLP}(x_{\mathrm{mlp}}),\\
x' &= \widehat{x} + g_{\mathrm{mlp}} \odot H_{\mathrm{mlp}},
\end{aligned}
\end{equation}
where $\mathrm{MLP}$ is a two-layer feed-forward network with
GELU activation and $g_{\mathrm{mlp}}$ is the MLP gate from
Eq.~\eqref{eq:app:wan-split}. The block output $x'$ has the same shape as the input $x$ and is passed to the next DiT block.

\subsection{Omni-Blocks}

Omni-blocks extend the dual-stream structure to simultaneously couple
video, text, and audio. Let
\begin{equation}
v \in \mathbb{R}^{B \times L_v \times D},
\quad
y \in \mathbb{R}^{B \times L_y \times D},
\quad
a \in \mathbb{R}^{B \times L_a \times D}
\end{equation}
be the current hidden states entering an omni-block.
We concatenate all three modalities:
\begin{equation}
\label{eq:app:omni-concat}
z = [\,v;\,y;\,a\,]
\in \mathbb{R}^{B \times (L_v + L_y + L_a) \times D}.
\end{equation}

\paragraph{Omni attention.}
Queries, keys, and values are constructed similarly to
Eq.~\eqref{eq:app:dit-qkv}, but RoPE is applied separately
to the video and audio spans (3D for video, 1D for audio),
and not to text:
\begin{equation}
\label{eq:app:omni-qkv}
\begin{split}
Q &= \mathrm{RoPE}_{\mathrm{v,a}}\big(z W_Q\big),\\
K &= \mathrm{RoPE}_{\mathrm{v,a}}\big(z W_K\big),\\
V &= z W_V.
\end{split}
\end{equation}
The attention output is
\begin{equation}
\label{eq:app:omni-attn}
H = \mathrm{softmax}\!\Big(
    \frac{Q K^\top}{\sqrt{D}}
\Big)V,
\qquad
O = H W_O,
\end{equation}
and we split $O$ into modality-specific spans
\begin{equation}
O = [\,O_v;\,O_y;\,O_a\,].
\end{equation}
Each span is passed through its own AdaLN-MLP branch, analogous
to Eq.~\eqref{eq:app:dual-update}, yielding
\begin{equation}
\label{eq:app:omni-update}
\begin{split}
v' &= v + g_{\mathrm{msa}}^{(\mathrm{v})} \odot O_v
       + g_{\mathrm{mlp}}^{(\mathrm{v})} \odot \mathrm{MLP}_v(\widehat{v}_{\mathrm{mlp}}),\\
y' &= y + g_{\mathrm{msa}}^{(\mathrm{y})} \odot O_y
       + g_{\mathrm{mlp}}^{(\mathrm{y})} \odot \mathrm{MLP}_y(\widehat{y}_{\mathrm{mlp}}),\\
a' &= a + g_{\mathrm{msa}}^{(\mathrm{a})} \odot O_a
       + g_{\mathrm{mlp}}^{(\mathrm{a})} \odot \mathrm{MLP}_a(\widehat{a}_{\mathrm{mlp}}).
\end{split}
\end{equation}

\subsection{Conditioning and Masking Strategies}

\paragraph{Caption dropout.}
During training, we randomly drop caption information with
probability $p_{\mathrm{cap}}$ by replacing the text embedding
with the unconditional caption embedding for a subset of samples.
This follows the usual classifier-free guidance~\cite{cfg} scheme and is
controlled by a binary mask $m_{\mathrm{cap}} \in \{0,1\}^B$:
\begin{equation}
y^{(0)}_b =
\begin{cases}
y_{\mathrm{cond},b}, & m_{\mathrm{cap},b} = 0,\\
y_{\mathrm{uncond},b}, & m_{\mathrm{cap},b} = 1.
\end{cases}
\end{equation}

\paragraph{Modality masking and dropping.}
To improve robustness under missing modalities, we apply
a lightweight masking or dropping scheme in omni-blocks.
At step $s$, we draw a Bernoulli variable with probability
\begin{equation}
p_{\mathrm{mask}}(s) = \max\!\Big(0,\ 1 - \tfrac{s}{S_{\max}}\Big),
\end{equation}
where $S_{\max}$ is a preset decay horizon.
If masking is enabled (\texttt{mask} mode), we construct a binary
mask over the concatenated sequence
in Eq.~\eqref{eq:app:omni-concat} that zeros out either the
audio or video span (but never both) before attention.
If dropping is enabled (\texttt{drop} mode), we instead remove the
corresponding span from the attention computation, effectively
training the model to handle single-modality inputs.

\section{Detailed Experimental Setup}
\label{sec:app:exp}

\subsection{Data Processing}

\paragraph{Landscape.}
The Landscape dataset~\cite{landscape} contains 1{,}000 audio-video clips, of which 900 are used for training and 100 for testing. All videos have a native resolution of $512\times288$ and a duration of 10\,s at 10\,fps (100 frames). We trim each clip uniformly to 9.7\,s (97 frames). The spatial resolution is preserved throughout. As Landscape provides category labels but no textual descriptions, we adopt the disentangled audio and video captions released by BridgeDiT~\cite{bridgedit}; the two captions are concatenated to form the final prompt used in our experiments.

\paragraph{AVSync15.}
The AVSync15 dataset~\cite{avsync} covers diverse acoustic events with varying native resolutions and clip durations. We follow the official split, using 1{,}350 clips for training and 150 for evaluation. We maintain each video's native spatial resolution, applying only minimal center-cropping when required by the VAE encoder (\textit{i.e.}, height and width divisible by~8). Videos are resampled to 12\,fps, and the number of frames is capped at 121 to ensure compatibility with the VAE temporal compression, yielding clips of approximately 10\,s. Audio is cropped to match the processed video length. We also adopt captions from BridgeDiT~\cite{bridgedit}. We find that this trimming does not remove semantically important content, as relevant acoustic events appear early in most clips.

\subsection{T2V Base Models.}
For the Wan-based experiments, we use the official \texttt{Wan2.1-T2V-1.3B} checkpoint as the video backbone, which contains 30 video blocks. Its parameters are kept frozen throughout training, and only the audio blocks and omni-blocks are updated; the exact numbers of these blocks are specified in the main paper. For the SD3-based experiments, we consider two T2V backbone configurations. The first is a 2.5B backbone with 28 video blocks, for which we train the full tri-modal framework from scratch. In this setting, we attach 14 audio blocks and 4 omni-blocks. The second is an 11B backbone pretrained on large-scale proprietary data. For this backbone we attach 21 audio blocks and 4 omni-blocks; all video-backbone parameters remain frozen, and only the audio blocks and omni-blocks are trained.

\subsection{Experimental Configuration}

All models are trained under the flow-matching objective~\cite{flowmatching}, and inference is performed with classifier-free guidance~\cite{cfg}. For each setting, we try several common guidance scales (\textit{e.g.}, 2, 3, 5, 7) to adjust the strength of the text conditioning during generation.

\paragraph{Wan-Adapted experiments.}
For the Wan-adapted setting~\cite{wan}, the results reported in the main paper correspond to a checkpoint trained for 90 full repeats of the dataset (approximately 81k sample-level updates). Training uses a global batch size of~8, implemented as per-device batch size of~1 across 8 GPUs with \texttt{accelerate} for data-parallel training. The learning rate is fixed to $1\times10^{-4}$. All videos are processed at their native resolution of $512\times288$, without spatial augmentation, and the audio sample rate is set to 16\,kHz.

\paragraph{SD3-Adapted experiments.}
For the SD3-adapted setting, each model variant is trained for 20k steps with a global batch size of~32 (per-device batch size of~1 across 32 NPUs, so approximately 640k sample-level updates). We adopt a maximum learning rate of $5\times10^{-5}$ with linear warmup during the first 5k steps, and training is performed using PyTorch FSDP. Video resolutions vary according to the original dataset and are fed to the model without resizing beyond the minimal VAE requirements; the audio sample rate is also fixed at 16\,kHz.

\section{Evaluation Metrics and Computation}
\label{sec:app:metrics}

We report four families of metrics: (i) overall quality (visual and audio),
(ii) text consistency, (iii) audio-video semantic consistency,
and (iv) temporal synchrony.
Let $\mathcal{G}$ and $\mathcal{R}$ denote generated and real sets, respectively.
Let $\Phi_v,\Phi_a,\Phi_y$ be video, audio, and text encoders, and
$\mathrm{cos}(\bp,\bq)=\frac{\bp^\top\bq}{\|\bp\|\,\|\bq\|}$ the cosine similarity.
All parameter settings and implementation details for the metrics follow JavisBench~\cite{javisDiT} to ensure consistency with prior work.

\subsection{Overall Quality}

\paragraph{Fr\'echet Video Distance (FVD, lower is better).}
For each video $g\in\mathcal{G}$ and $r\in\mathcal{R}$, we extract
I3D features
$\bz_i=\Phi_v^{\mathrm{I3D}}(g_i)$ and
$\by_j=\Phi_v^{\mathrm{I3D}}(r_j)$.
Let $(\bmu_g,\bSigma_g)$ and $(\bmu_r,\bSigma_r)$ be their empirical mean and covariance:
\begin{equation}
\begin{aligned}
\mathrm{FVD}
&= \big\|\bmu_g - \bmu_r\big\|_2^2 \\
&\quad + \mathrm{Tr}\!\Big(
    \bSigma_g + \bSigma_r
    - 2\big(\bSigma_g^{1/2}\bSigma_r\bSigma_g^{1/2}\big)^{1/2}
\Big).
\end{aligned}
\end{equation}

\paragraph{Fr\'echet Audio Distance (FAD, lower is better).}
For audio, we extract AudioCLIP embeddings
$\ba_i=\Phi_a^{\mathrm{AC}}(g_i)$ and
$\bb_j=\Phi_a^{\mathrm{AC}}(r_j)$, with empirical Gaussians
$(\bmu_g^{(a)},\bSigma_g^{(a)})$ and $(\bmu_r^{(a)},\bSigma_r^{(a)})$:
\begin{equation}
\begin{aligned}
\mathrm{FAD}
&= \big\|\bmu_g^{(a)} - \bmu_r^{(a)}\big\|_2^2 \\
&\quad + \mathrm{Tr}\!\Big(
    \bSigma_g^{(a)} + \bSigma_r^{(a)}
    - 2\big(\bSigma_g^{(a)\,1/2}\bSigma_r^{(a)}\bSigma_g^{(a)\,1/2}\big)^{1/2}
\Big).
\end{aligned}
\end{equation}

\subsection{Text Consistency}

Let $y_g$ be the text prompt for sample $g\in\mathcal{G}$, and
$\bt_g = \Phi_y(y_g)$ its text embedding.


\paragraph{CLIP-Score (text-to-video, higher is better).}
For CLIP encoders $(\Psi_v^{\mathrm{CLIP}},\Psi_y^{\mathrm{CLIP}})$,
we uniformly sample $F_g=48$ frames from each video $v_g$ (using linear spacing) and compute frame embeddings
$\mathbf{f}_{g,k}=\Psi_v^{\mathrm{CLIP}}(v_{g,k})$ and
a text embedding
$\mathbf{t}_g^{\mathrm{CLIP}}=\Psi_y^{\mathrm{CLIP}}(y_g)$.
After $\ell_2$-normalization, the per-sample CLIP score is
\begin{equation}
\mathrm{CLIPScore}(g)
= \frac{1}{48}\sum_{k=1}^{48}
  \mathrm{cos}\!\big(\mathbf{f}_{g,k},\ \mathbf{t}_g^{\mathrm{CLIP}}\big),
\end{equation}
and the dataset-level score is
\begin{equation}
\mathrm{CLIPScore}
= \frac{1}{|\mathcal{G}|}
  \sum_{g\in\mathcal{G}}
  \mathrm{CLIPScore}(g).
\end{equation}

\paragraph{CLAP-Score (text-to-audio, higher is better).}
For CLAP encoders $(\Psi_a^{\mathrm{CLAP}},\Psi_y^{\mathrm{CLAP}})$,
with audio $a_g$ resampled to $48$\,kHz as required by CLAP, we define
\begin{equation}
\mathrm{CLAPScore}
= \frac{1}{|\mathcal{G}|}
  \sum_{g\in\mathcal{G}}
  \mathrm{cos}\!\big(
    \Psi_y^{\mathrm{CLAP}}(y_g),\
    \Psi_a^{\mathrm{CLAP}}(a_g)
  \big).
\end{equation}

\subsection{Audio-Video Semantic Consistency}

\paragraph{Global ImageBind AV (IB-AV, higher is better).}
Using ImageBind video/audio encoders
$\Phi_v^{\mathrm{IB}}$ and $\Phi_a^{\mathrm{IB}}$, we obtain
global embeddings $\mathbf{v}_g^{\mathrm{IB}}$ and
$\mathbf{a}_g^{\mathrm{IB}}$ for each sample $g$ and define
\begin{equation}
\mathrm{IB\text{-}AV}
= \frac{1}{|\mathcal{G}|}
  \sum_{g\in\mathcal{G}}
  \mathrm{cos}\!\big(
    \mathbf{v}_g^{\mathrm{IB}},\ \mathbf{a}_g^{\mathrm{IB}}
  \big).
\end{equation}

\paragraph{CAVP-score (higher is better).}
We use a pretrained CAVP model to extract video features $\mathbf{v}_g^{\mathrm{CAVP}}$ and audio features $\mathbf{a}_g^{\mathrm{CAVP}}$. For audio $a_g$: (i) resample to 16\,kHz; (ii) convert to a Mel-spectrogram via $\text{get\_spectrogram}(\cdot)$; (iii) process via CAVP's spectrogram encoder and projection head. For video $v_g$, we use 4 FPS following JavisBench~\cite{javisDiT}. The CAVP score is:
\begin{equation}
\mathrm{CAVP}
= \frac{1}{|\mathcal{G}|}
  \sum_{g\in\mathcal{G}}
  \mathrm{cos}\!\big(
    \mathbf{v}_g^{\mathrm{CAVP}},\ \mathbf{a}_g^{\mathrm{CAVP}}
  \big).
\end{equation}

\paragraph{AVHScore (higher is better).}
For AVHScore, we compute ImageBind frame embeddings
$\mathbf{v}_{g,t}^{\mathrm{IB}}$ for each frame $t$ and a global
audio embedding $\mathbf{a}_g^{\mathrm{IB}}$:
\begin{equation}
\begin{aligned}
\mathrm{AVHScore}(g)
&= \frac{1}{T_g}
   \sum_{t=1}^{T_g}
   \mathrm{cos}\!\big(
     \mathbf{v}_{g,t}^{\mathrm{IB}},\ \mathbf{a}_g^{\mathrm{IB}}
   \big), \\
\mathrm{AVHScore}
&= \frac{1}{|\mathcal{G}|}
   \sum_{g\in\mathcal{G}}
   \mathrm{AVHScore}(g).
\end{aligned}
\end{equation}

\begin{figure*}[htbp]
  \centering
  \includegraphics[width=1\textwidth]{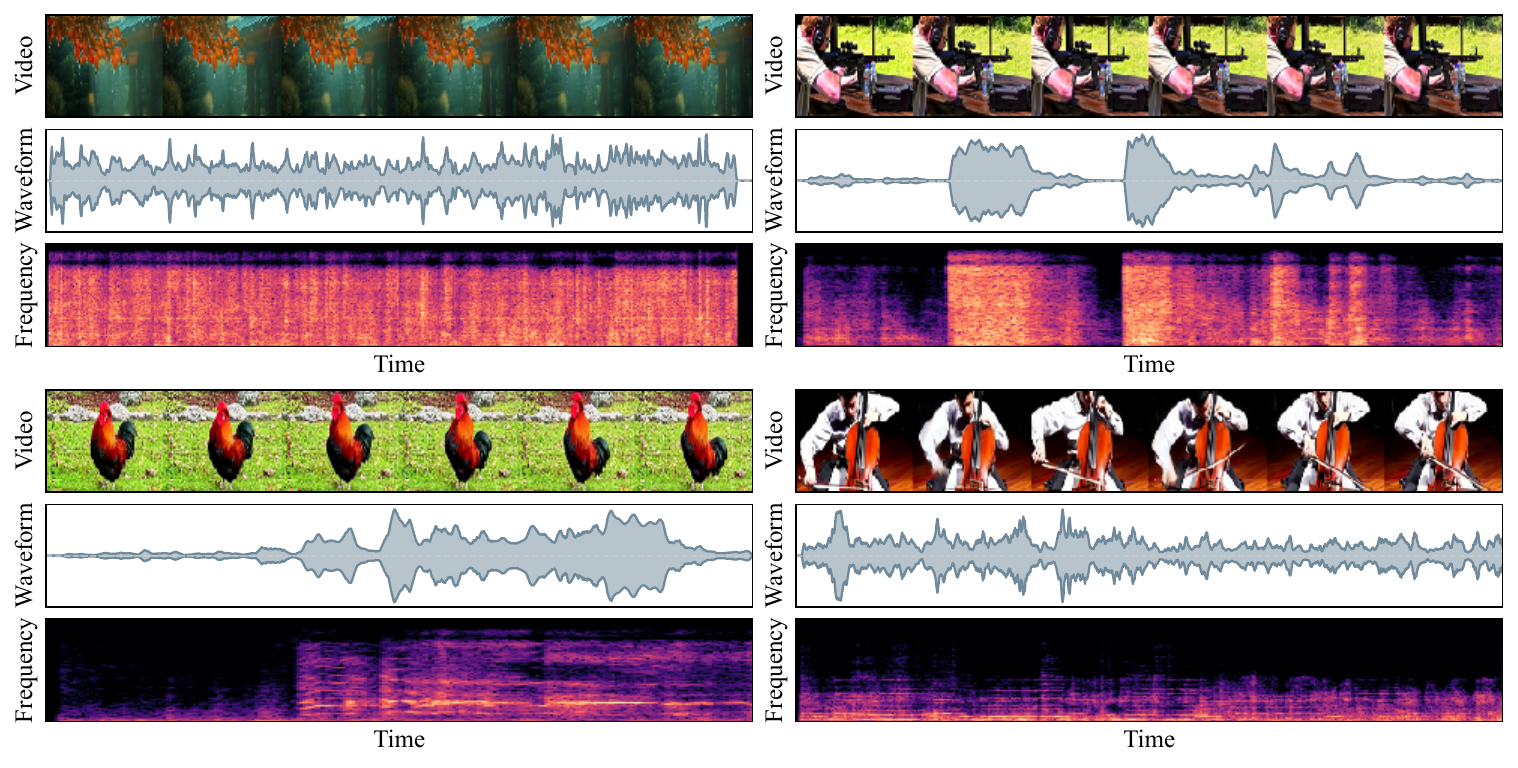}
  \caption{
  \textbf{Case Studies.}
  The top-left example is generated by the Wan-adapted model trained and evaluated on the Landscape dataset, while the remaining three examples are produced by the SD3-adapted models trained and evaluated on AVSync15.
  Each panel shows uniformly sampled video frames (six per sample), the corresponding waveform, and an 0-10\,kHz log-mel spectrogram.
  The four cases illustrate distinct acoustic events: rain, gunshots, rooster crowing, and cello play (ordered left-to-right, top-to-bottom).
  }
  \vspace{-2pt}
  \label{fig:case_app}
\end{figure*}

\subsection{Temporal Synchrony}

\paragraph{AV-Align (higher is better).}
AV-Align measures the temporal synchrony between audio energy peaks and video motion peaks using Intersection-over-Union (IoU). For each sample $g\in\mathcal{G}$:
\begin{enumerate}
    \item Audio Peak Detection: Extract audio energy peaks $P_a = \{t_{a,1}, t_{a,2}, ..., t_{a,K}\}$ (timestamp set) via audio energy analysis.
    \item Video Peak Detection: Extract video frames, compute optical flow trajectories to measure motion intensity, and detect motion peaks $P_v = \{t_{v,1}, t_{v,2}, ..., t_{v,L}\}$ (timestamp set).
    \item IoU Calculation: Convert timestamps to frame indices using the video FPS, then compute the IoU of $P_a$ and $P_v$:
    \begin{equation}
    \mathrm{AV\text{-}Align}(g) = \frac{|P_a \cap P_v|}{|P_a \cup P_v|},
    \end{equation}
    where $|\cdot|$ denotes the cardinality of the set.
\end{enumerate}
The dataset-level AV-Align is the mean over all samples $g\in\mathcal{G}$:
\begin{equation}
\mathrm{AV\text{-}Align} = \frac{1}{|\mathcal{G}|} \sum_{g\in\mathcal{G}} \mathrm{AV\text{-}Align}(g).
\end{equation}

\paragraph{JavisScore (higher is better) proposed by JavisBench.}
This metric evaluates audio-video temporal synchronization by partitioning video and audio into overlapping windows. Its core idea is to compute frame-wise similarity between visual embeddings of video window frames and audio embeddings of corresponding audio windows (both extracted via ImageBind), then average the bottom fraction of similarity scores per window to emphasize synchronization robustness. All parameter settings follow JavisBench~\cite{javisDiT} for consistency with prior work.

\section{Case Study}

We provide additional qualitative examples in Fig.~\ref{fig:case_app}. Across both backbone variants and datasets, the adapted models exhibit good synchrony and consistently produce promising generation results.

\section{Discussion and Future Directions}
\label{sec:app:discussion}
While our framework demonstrates encouraging performance, several limitations remain and point toward promising avenues for future work. First, even when reusing a pretrained T2V backbone, the introduction of audio blocks and omni-blocks substantially increases the number of trainable layers. Achieving strong performance therefore requires extensive optimization, and the omni-blocks in particular must still be trained from scratch. This poses non-trivial demands on both dataset size and computational resources, especially when compared with dual-backbone adaptation strategies explored in prior work. Nonetheless, we believe that an end-to-end trainable paradigm remains fundamentally more promising. In contrast, adapting to heterogeneous pretrained backbones often introduces significant engineering overhead and limits generalizability across architectures.

Second, fine-grained tri-modal fusion depends not only on the quality of audio–video training data but also heavily on the reliability of textual captions. Misalignment between the caption and the underlying audio–video content can hinder model training and degrade synchronization performance. In our experiments, we observed that existing open-source caption-generation pipelines—such as the HVGC framework~\cite{bridgedit}, which relies on VLMs and LLMs with prompt engineering—often infer audio descriptions solely from the visual modality. This introduces an intrinsic modality mismatch and may inadvertently propagate audio–video desynchronization into the supervision signal. Future work will explore more compute-efficient training strategies, such as parameter-efficient tri-modal adaptation or layer-sharing schemes across modalities, as well as improved approaches for constructing and leveraging text captions. In particular, jointly grounded captioning pipelines that integrate audio, video, and language signals hold strong potential for reducing annotation noise and strengthening tri-modal alignment.

Finally, reliable and universally accepted metrics for audio–video synchrony remain lacking. Recent analyses from JavisBench~\cite{javisDiT}, for example, suggest that AV-Align~\cite{avalign} can be unstable or misleading under complex motion patterns, overlapping acoustic events, or non-stationary backgrounds. This limitation is especially critical for tri-modal generation: without a trustworthy synchrony metric, it is difficult to systematically compare models, guide architectural choices, or perform principled hyperparameter tuning. Developing principled quantitative metrics that faithfully capture perceptual audio–video synchronization is therefore an important direction for future research.


\end{document}